\documentclass[manuscript,screen]{acmart}

\AtBeginDocument{%
  \providecommand\BibTeX{{%
    \normalfont B\kern-0.5em{\scshape i\kern-0.25em b}\kern-0.8em\TeX}}}

\usepackage{colortbl}
\usepackage{multirow}

\usepackage{microtype}        
\usepackage[all]{hypcap}    
\usepackage{ccicons}          

\definecolor{author1}{rgb}     {0.9,0.5,0.0}
\definecolor{author2}{rgb}     {0.82, 0.1, 0.26}
\definecolor{fixme}{rgb}     {0.42, 0.6, 0.26}

\begin{document}

\title{Investigating Performance and Usage of Input Methods for Soft Keyboard Hotkeys}


\author{Katherine Fennedy}
\affiliation{%
  \institution{Singapore University of Technology and Design (SUTD)}}
\email{katherine_fennedy@mymail.sutd.edu.sg}

\author{Sylvain Malacria}
\affiliation{%
  \institution{Inria Lille - Nord Europe}
  \country{France}}
\email{sylvain.malacria@inria.fr}

\author{Hyowon Lee}
\affiliation{%
  \institution{Dublin City University}
  \country{Ireland}}
\email{hyowon.lee@dcu.ie}

\author{Simon T. Perrault}
\affiliation{%
  \institution{Singapore University of Technology and Design (SUTD)}}
\email{perrault.simon@gmail.com}

\renewcommand{\shortauthors}{Fennedy et al.}

\begin{abstract}
Touch-based devices, despite their mainstream availability, do not support a unified and efficient command selection mechanism, available on every platform and application.
We advocate that hotkeys, conventionally used as a shortcut mechanism on desktop computers, could be generalized as a command selection mechanism for touch-based devices, even for keyboard-less applications.
In this paper, we investigate the performance and usage of soft keyboard shortcuts or hotkeys (abbreviated \emph{SoftCuts}) through two studies comparing different input methods across sitting, standing and walking conditions.
Our results suggest that SoftCuts not only are appreciated by participants but also support rapid command selection with different devices and hand configurations. We also did not find evidence that walking deters their performance when using the Once input method.
\end{abstract}



\maketitle

\begin{figure}[h]
\centering
  \includegraphics[width=1.0\columnwidth]{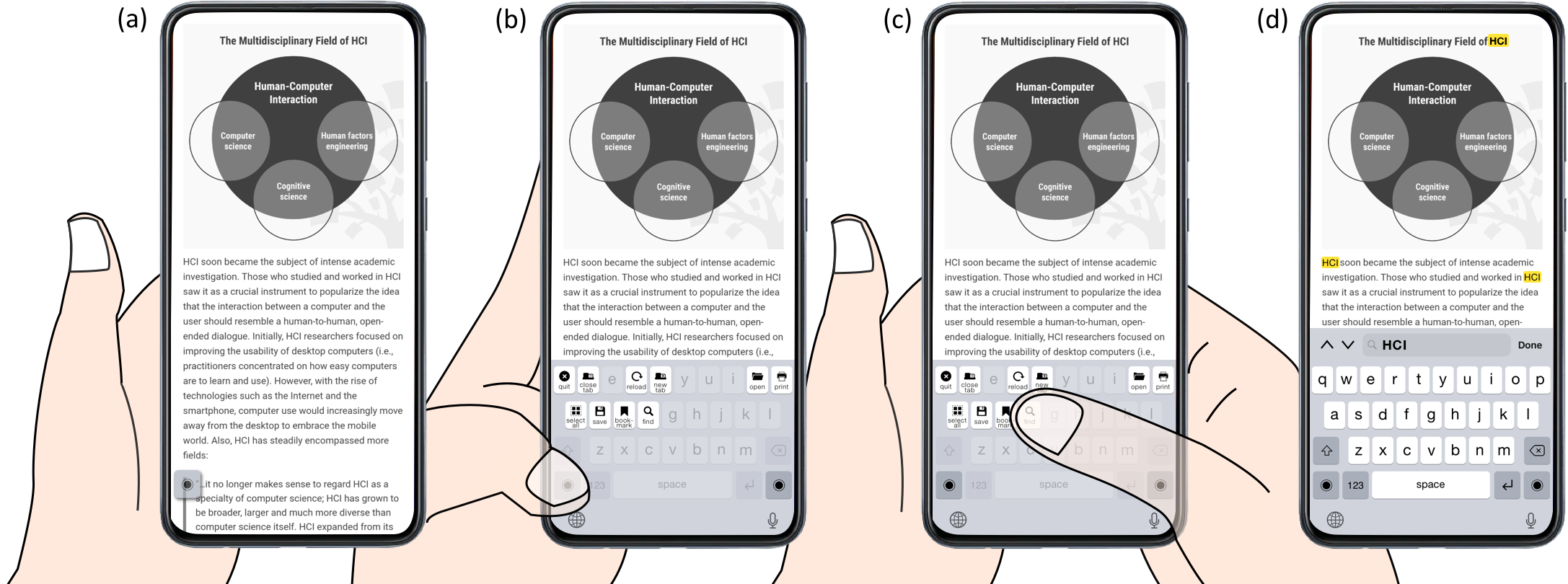}
  \caption{Web browsing scenario of soft keyboard shortcuts using \emph{Once} method. (a) a semi-transparent modifier key is available in the browser, (b) when user taps it, all available shortcuts will be presented to facilitate browsing, (c) user then taps on the "F" key (find command) and hence, (d) every occurrence of the ``HCI'' word (as typed by the user) is highlighted in yellow.}~\label{fig:banner}
\end{figure}

\section{Introduction}
Hotkeys (also known as keyboard shortcuts) are a well-established shortcut mechanism available on every desktop keyboards that allows users to select most frequent commands rapidly \cite{Malacria2013_ExposeHK}.
Touch-based devices, on the other hand, do not offer such a unique and unified shortcut mechanism, available on every platform and application.
This may be one reason why, despite the mainstream availability, touch-based devices still cannot compete with desktop operating systems (OS) for many productivity tasks \cite{Tero2015}.

Physical keyboards became an exception rather than the norm on tablets and smartphones, where they are usually replaced with soft keyboards (software keyboards) that are made available only when text entry is required.
This may be why hotkeys never seem to have been considered as a viable command selection standard for these devices.
Another possible reason is that the primary strategies used to convey hotkeys in the desktop OS are tooltips (when the cursor hovers a widget) and labels (next to the menu item in the menu bar), methods that are not readily adaptable to the touch-based OS.

Attempts to instantiate them can be found on some commercial touch-based OS like the latest Microsoft Surface \cite{Surface} and Samsung \cite{Samsung} tablets. A subset of available commands can be selected using keyboard shortcuts on the soft keyboard, by holding/tapping a modifier (Control/Command) key and subsequently hitting the key associated with a command.
Another, yet different, example could be found in previous implementations of the Swype keyboard \cite{Nuance2017} where gestures could be performed by swiping from a dedicated button to specific keys to activate commands.
However, commands associated with keys are not exposed to the user who has either to guess them, discover them through repeated tries, or through online resources like websites. 
Regardless, these mechanisms are limited to scenarios where the keyboard is displayed and constrained to one specific input method (taps or swipes).
More surprisingly, these commercial solutions do not leverage the dynamic nature of soft keyboards' layout that could be changed at will, making the discovery and learning of hotkeys even easier than on desktop systems. In summary, the motivation to further study hotkey-based command selection is driven by the following reasons:
\begin{itemize}
  \item Hotkeys on touchscreens already exist, albeit not standardised in terms of input method and visual presentation.
  \item It has the potential to become a multi-device command mechanism, available on desktops and touch devices, where advanced users can reuse pre-existing knowledge from using desktop computers or the other way around. 
  \item The keyboard layout provides a familiar canvas to maintain spatial consistency across applications for a given command.
\end{itemize}

In this paper, we advocate for the generalization of soft keyboard shortcuts or hotkeys (abbreviated \emph{SoftCuts}) as a viable command selection mechanism on touch-based devices, and focus on investigating their performance and usage through two studies.
Our first study compares the performance of three different input methods (\emph{Once}, \emph{Swipe} and \emph{User Maintained}) for SoftCuts across different devices, orientations and number of hands used to interact regardless of the visual presentation.
Our second study evaluates the usage of the same three input methods across varied mobility conditions like sitting, standing and walking.
Our results suggest that while the input method \emph{Once} (based on two successive taps on soft keys) overall performed best, some users also like using \emph{Swipe} for one-handed interaction as well as \emph{User Maintained} for Tablet.
Altogether, these results confirm that SoftCuts could be generalized as an efficient command selection mechanism for touch-based devices.

\section{Related Work}
\subsection{Soft Keyboard Hotkeys in HCI}
Currently, soft keyboard hotkeys have been successfully implemented by Samsung and Microsoft. We discuss these implementations in the next section. However, to the best of our knowledge, no academic work has been published on the topic. With a vast design space at our disposal, we decided to focus on input performance first, as it is one of the most important criteria considered in command selection literature.

\subsection{Command Selection with Keyboard Shortcuts}
Command selection on desktop computers mostly relies on pointing with the cursor on commands organized in menubars and toolbars.
However, alternative methods using keyboards have been proposed.
This is typically the case of Alt key navigation (part of the accessibility interface in Microsoft Office) that displays keys associated with successive levels of the menu hierarchy when the `Alt` key is pressed, allowing users to select commands using the keyboard.
However, it has been shown that users had difficulties in chunking these key sequences into one single cognitive unit \cite{Miller2011}.
Another alternative is keyboard shortcuts (also called hotkeys) that allow a wide range of commands to be selected with a single key combination, thus removing the need to traverse the hierarchy.
Repeated studies have demonstrated the performance benefits of hotkeys over alternative input methods such as menubars or toolbars \cite{Card1980, Lane2005,Miller2011, Odell2004}.
The high performance ceiling of hotkeys has motivated researchers to investigate how they could leverage different keyboard capabilities \cite{Bailly2013}, hand postures \cite{Zheng2018_FingerArc} or different types of key sequences \cite{Au2016}. 
Nevertheless, hotkeys still suffer from one main limitation, which is that users must \textit{recall} the key combination in order to be able to use them~\cite{Grossman2007,Malacria2013_Skillometer}.
As a result, in spite of their performance benefits, hotkeys remain limited in use~\cite{Bhavnani2000,Lane2005}.
Hardware solutions can be used to overcome the recall necessity of hotkeys. Typically, Optimus keyboards~\cite{optimus} are physical keyboards whose each key is a display that can dynamically change, typically to reveal the commands associated when a modifier key is pressed, hence making interaction with hotkeys rely on \textit{recognition} rather than \textit{recall}.
Similar software solutions have also been designed so that users can expose the hotkeys on the monitor while pressing a modifier key \cite{Giannisakis2017,Malacria2013_ExposeHK,Tak2013satisficing} and demonstrated the efficiency of hotkeys even when users are not aware of the key combination beforehand.

\subsection{Command Selection with Touch-based Devices}
Similar to desktop computers and despite of supporting multitouch input capabilities, command selection on commercial touch-based devices still mostly rely on single point tap-based interaction, hence requiring selection commands to be located in toolbars or hamburger menus\footnote{\url{https://en.wikipedia.org/wiki/Hamburger_button}}.
Exceptions will be found with \textit{Swhidgets} buttons, hidden by default and that users first uncover with a swipe gesture relying on the metaphor of sliding physical objects \cite{PONGISS19, Schramm2016}, or in device shaking gestures such as the shake-to-undo in iOS. 
Researchers though have long explored how different modalities, such as gestural interaction, could be adapted to facilitate command selection on touch-based devices.
The widely studied marking menus \cite{Kurtenbach1994} have influenced many stroke-based strategies \cite{Appert2009,Francone2009,Francone2010,Kin2011,Lepinski2010,Zheng2018_M3} to trigger commands on multitouch platforms.
Similarly, MelodicTap \cite{Heo2016} and Arp\`{e}ge \cite{Ghomi2013} investigated rhythmic chorded gestures to expand input vocabulary on touch screens.
However, as Norman pointed out, gestural interaction can be ``a step backward in usability'' \cite{Norman2010_B} for reasons like poor discoverability and inconsistent mapping.
The unnatural \cite{Norman2010_A} and invisible mapping between a gestural input and the corresponding command output is why users need to learn and rehearse before being able to invoke such shortcuts on-demand \cite{Avery2016}.
FastTap \cite{Gutwin2014} and its adaptation to smartwatches \cite{Lafreniere2016} alleviate this problem by relying on a full-screen static grid of commands that can be triggered through a chorded selection.
Compared to SoftCuts, FastTap relies on a full-screen grid layout, supports only UM, and is rehearsal-based, offers novice and expert modes of interaction, displaying the menu only after a delay. 
However, several studies have suggested that users have difficulties to adopt expert mode~\cite{Lafreniere2017, Goguey2019} and also investigated the necessity of the delay-based appearance~\cite{Henderson2020} of Marking Menus, hence raising similar questions for existing techniques, such as FastTap.

Command selection techniques leveraging soft keyboard have also been explored in the literature.
Initially designed for ``pen-based'' devices, one early example can be found in Command Strokes \cite{Kristensson2007} where users can select a command by drawing its name on a soft keyboard, starting from the Ctrl key, as they would with a gestural keyboard \cite{Kristensson2004}.
More recently, Alvina et al. presented CommandBoard \cite{Alvina2017}, which combines gestural text entry with an Octopocus menu \cite{Bau2008} to merge typing and command execution into one continuous user action. 
Novice users can pause after having written the word to display the gestural menu and select the desired command.
Ando et al. \cite{Ando2019} proposed a different technique merging text selection with command execution, by starting a drag gesture from a specific key of the keyboard. The first key specifies the command while the drag gesture is used to select the desired text.
However, this technique is limited to text associated commands and does not teach which command is associated with each key.

\vspace{3mm}
As a summary, while various command selection mechanisms for touch-based devices have been proposed, some leveraging the keyboard area, the performance and usage of hotkeys on soft keyboards have not been investigated yet. They would have the advantage of being easy to discover (with an additional key on the soft keyboard layout), and we could leverage users' familiarity with that mechanism as it is already widely implemented on desktop and laptop computers using physical keyboards.

\section{SoftCuts}
As seen in previous sections, it is surprisingly rare for soft keyboards to support the activation of hotkeys. However, we believe that soft keyboard shortcuts/hotkeys (\emph{SoftCuts}) could easily be generalized for every touch-based device, not only as a shortcut mechanism, but as a generic command selection mechanism. Since soft keyboards are displayed on a screen, they are dynamic in nature and can adapt their layout and appearance.
For example, the keyboard layout would change to display capital letters when the ``Shift'' key is pressed.
It could similarly be adapted when a modifier key (e.g. Ctrl or Cmd) is pressed to display its associated command on each key, therefore helping users to discover the commands.

\subsection{Existing Implementations}
\begin{figure}
    \centering
     \includegraphics[width=1.0\columnwidth]{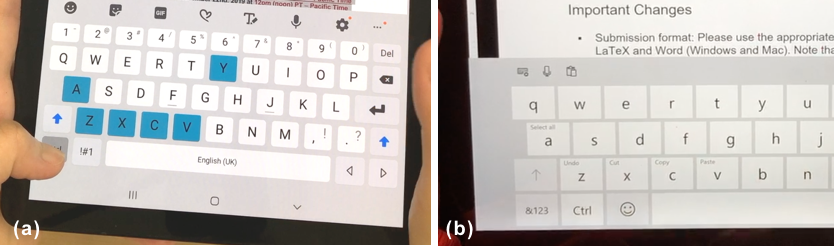}
    \caption{Existing Implementations of Soft Keyboard Hotkeys on (a) Samsung Galaxy S, with the hotkeys highlighted in blue once the user presses the Ctrl key, (b) Microsoft Surface, with the name of the commands associated with hotkeys displayed after the user taps on the Ctrl key. Note: Samsung uses User Maintained as an input method, while Microsoft uses Once.}~\label{fig:existing-skh}
\end{figure}

Latest Microsoft Surface \cite{Surface} and Samsung Galaxy S \cite{Samsung} tablets support the selection of a subset of available commands using keyboard shortcuts (Figure~\ref{fig:existing-skh}), by respectively hitting or holding a modifier key (Ctrl) and then hitting a key associated with a command.
However, commands are poorly suggested to the user: in Microsoft implementations, the command is scarcely displayed using only text over the associated key, where on Samsung tablets, keys associated with commands are only highlighted without describing which command would be selected.
As a result, these commercial solutions do not fully leverage the dynamic nature of soft keyboards that may make the discovery and learning of hotkeys even easier than on desktop systems, as the keyboard layout could be changed at will. 
Moreover, these mechanisms are limited to scenarios where the keyboard is displayed and constrained to one specific input method.

\subsection{Input Methods}
Unlike desktop hotkeys, SoftCuts do not require users to memorize the hotkeys beforehand, but leverage spatial memory, making them a viable command selection technique.
SoftCuts also allow three different input methods to coexist:

\begin{enumerate}
    \item \emph{User Maintained interaction (UM)}: Holding the modifier key with one finger and activating the hotkey with a different finger. This mechanism is similar to hotkeys on desktop computers and is also used on Samsung Galaxy S tablets.
    \item \emph{Two sequential taps (Once)}: Tapping the modifier key (without holding) changes the layout of the keyboard to display commands that can then be selected. Subsequent tapping on any key will change the layout back to QWERTY automatically (see Figure~\ref{fig:banner}). This mechanism is similar to how the Shift key behaves to capitalize text with soft keyboards, and to how hotkeys are implemented in recent Microsoft Surface tablets.
    \item \emph{Sliding gestures (Swipe)}: Touching the modifier changes the layout of the keyboard to display the commands, that can be selected by sliding the finger over the key and releasing it.
\end{enumerate}

\begin{figure}
    \centering
     \includegraphics[width=0.6\columnwidth]{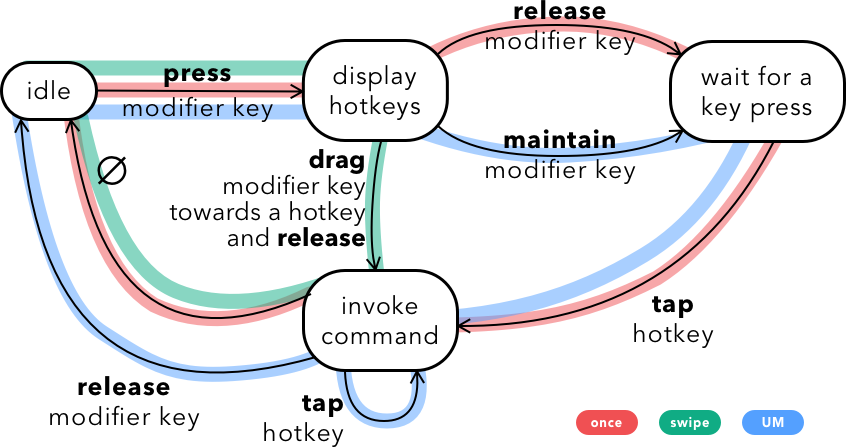}
    \caption{State diagrams for each \emph{Input Method}, showing that the methods can be used concurrently on a system. Note: \emph{UM} allows user to perform multiple command selections by tapping multiple hotkeys sequentially.}~\label{fig:expe5-states}
\end{figure}

We illustrate how these three input methods are compatible with each other, in the state diagram (see Figure~\ref{fig:expe5-states}), leaving users the freedom to choose their preferred method depending on the context. 

In the following sections, we will elaborate on two studies investigating the performance and usage of each input method. In the first experiment, we investigate the performance of each method and see which one(s) the user prefers. While a previous study~\cite{ElBatran2014} theoretically estimated that a swipe is approximately 10ms faster than a single pointing gesture, it is unclear how that result would translate to our unique context of keyboard layout and two sequential taps instead of a single one. In the second experiment, we investigate how different level of physical activity may impact the usage frequency. Which method performs the best? Do people switch methods as they wish, or is there a particular one they would prefer? How will the results change in different mobility conditions? The answers will be revealed next.

\section{Experiment 1: Performance of Input Methods}
In this study, we compare the performance of the three SoftCuts' input methods (\emph{UM}, \emph{Once} and \emph{Swipe}) in terms of time, accuracy and subjective preference.
Since we envision SoftCuts as a general command selection mechanism for touch-based devices, we decided to test these input methods in various configurations, with both mobile phones and tablets, in landscape and portrait orientations and with one and two hands to perform the interaction.

\subsection{Participants and Apparatus}
Twelve participants (all right-handed, 5 female), aged 18 to 30 ($M=25.1$, $SD=3.29$) were recruited from the university community. They received the equivalent of 7.5 USD for their participation.
One participant used a tablet every day while three used it at least once per month. Only one participant was familiar with swipe-based keyboards.
The experimental software was written in Java using Android Studio as an IDE and ran on version 28 of Android SDK. We used a Samsung A10 phone (6.2'', 168 grams) running Android 9 for the Phone condition and a Samsung Galaxy Tab 4 (10.5'' screen, 483 grams) running Android 9 for the Tablet condition.

\subsection{Procedure, Task and Stimulus}
Participants began the experiment with a questionnaire that asked about their demographic information, familiarity with tablet and use of swipe-based keyboards.
They were then briefed on the experiment and were presented the three \emph{Input Methods} by the experimenter.
They would then be asked to perform repeated series of command executions as rapidly and accurately as possible, with a given \emph{Input Method} and for a given \emph{Configuration}.
A \emph{Configuration} corresponds to the combination of a \emph{Device} (Phone or Tablet), a given \emph{Orientation} (Portrait or Landscape) and \emph{Handedness} (one or two hands).
The \emph{Configurations} compared in this study are detailed below.

\begin{figure}[h]
\centering
  \includegraphics[width=1.0\columnwidth]{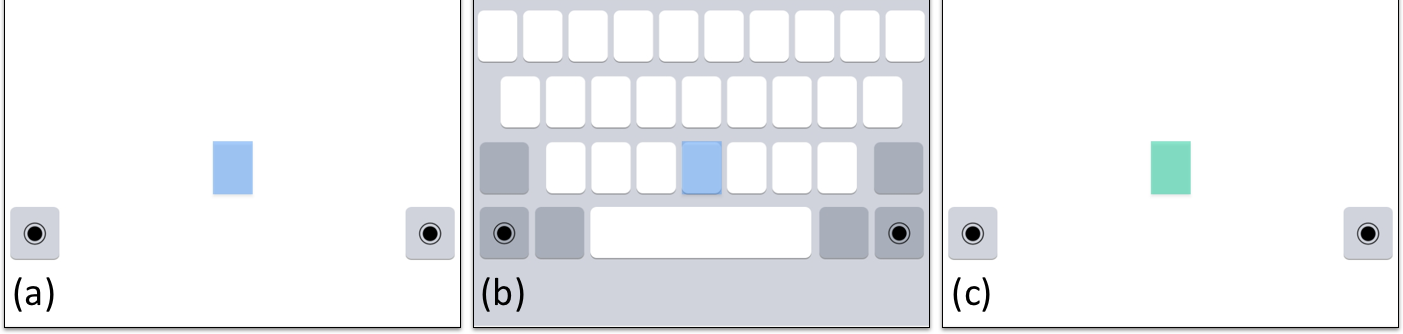}
  \caption{Example of a trial on smartphone. (a) The stimulus is highlighted in blue at the beginning of the trial to minimize visual search (b) the user pressed on one of the modifier keys, (c) upon correct selection, the key would turn green (red if incorrect).}~\label{fig:exp2task}
\end{figure}

As we were interested in the performance of command execution and wanted to eliminate other factors such as visual search, each trial began by specifying the target command by simply highlighting its key in blue (Figure~\ref{fig:exp2task}(a)). By doing so, we also simulate expert users~\cite{Bailly2016}.
Participants then had to select the target command using the requested \emph{Input Method}.
In that respect, participants could select the commands using either of the two modifier keys, respectively located on the bottom left and right corners of the keyboard (Figure~\ref{fig:exp2task}).
After performing the selection, feedback would appear on the screen for 500 milliseconds with the target being highlighted in either green (if correct) or red (if incorrect), before proceeding to the next trial (Figure~\ref{fig:exp2task}(c)).
A trial was considered successful if the participant was able to select the correct target.
Trial completion time was measured from target appearance to target selection.

There were 6 possible targets (Q, Z, T, V, P and M) as these keys are spread around the keyboard.
For each condition (\emph{Device}$\times$\emph{Orientation}$\times$\emph{Handedness} combination), participants repeated 1 training block of 6 trials (once each target, in random order), followed by 3 test blocks of 12 trials (twice each target, in random order).
After each condition, participants would provide subjective feedback on a 1-5 Likert Scale (1: strongly disagree, 5: strongly agree) on their overall opinion for the method ("I liked using this input method"), ease of use, comfort, perceived speed and perceived accuracy.
Participants then proceeded to the next condition for that part.

\subsection{Design}

\subsubsection{Experimental Conditions}
Our initial design space for this experiment includes: Input Methods \{UM, Once, Swipe\}, Device \{Phone, Tablet\}, Handedness \{One, Two\}, Orientation \{Landscape, Portrait\}.
Complete factorial exploration of this design space would lead to a total of 3 $\times$ 2 $\times$ 2 $\times$ 2 = 24 combinations.
However, we decided to discard 10 combinations that we considered as impractical or impossible, to focus our study on 14 combinations that were considered comfortable and/or doable (see Table~\ref{tab:expe2-conditions}).
The experiment was split in two parts (one- and two-handed).
The order of appearance of both parts was fully counter balanced across participants.
\begin{table}[]
\small
\centering
\begin{tabular}{ccc|ccc}
\textbf{Hands} & \textbf{Device} & \textbf{Orientation} & \begin{tabular}[c]{@{}c@{}}\textbf{User}\\ \textbf{Maintained}\end{tabular} & \textbf{Once} & \textbf{Swipe} \\ \hline
 &  & Portrait & \cellcolor[HTML]{C0C0C0} & \multicolumn{2}{c}{\cellcolor[HTML]{FFFFC7}1-handed part} \\
 & \multirow{-2}{*}{Phone} & Landscape & \cellcolor[HTML]{C0C0C0} & \multicolumn{2}{c}{\cellcolor[HTML]{C0C0C0}} \\
 &  & Portrait & \cellcolor[HTML]{C0C0C0} & \multicolumn{2}{c}{\cellcolor[HTML]{C0C0C0}} \\
\multirow{-4}{*}{1} & \multirow{-2}{*}{Tablet} & Landscape & \multirow{-4}{*}{\cellcolor[HTML]{C0C0C0}\begin{tabular}[c]{@{}c@{}}2\\ Hands\\ Required\end{tabular}} & \multicolumn{2}{c}{\multirow{-3}{*}{\cellcolor[HTML]{C0C0C0}\begin{tabular}[c]{@{}c@{}}Physical\\ Constraints\end{tabular}}} \\ \hline
 &  & Portrait & \multicolumn{3}{c}{\cellcolor[HTML]{FFFFC7}} \\
 & \multirow{-2}{*}{Phone} & Landscape & \multicolumn{3}{c}{\cellcolor[HTML]{FFFFC7}} \\
 &  & Portrait & \multicolumn{3}{c}{\cellcolor[HTML]{FFFFC7}} \\
\multirow{-4}{*}{2} & \multirow{-2}{*}{Tablet} & Landscape & \multicolumn{3}{c}{\multirow{-4}{*}{\cellcolor[HTML]{FFFFC7}2-handed part}}
\end{tabular}
\caption{\label{tab:expe2-conditions}Experimental Conditions for Study 1. Greyed-out conditions were not tested: UM could not be tested in 1-handed conditions as it requires two hands; During pre-tests, participants were not able to reach every key in the 1-handed Phone-Landscape conditions}
\end{table}

\subsubsection{One-Handed Part}
For the one-handed part, participants were invited to hold, with their preferred hand, a Phone in Portrait orientation only.
Tablet device and Landscape orientations were discarded either for physical constrains or because two hands were required for comfortable interaction.
Similarly, \emph{UM} was excluded as it requires two simultaneous contact points which cannot be achieved in these configurations.
This results in a within-subject design with \emph{Input Method} as a two-level \{ Once, Swipe \} independent variable, fully counterbalanced across participants.
In total, a participant would perform 2 Input Methods $\times [($1 training block $\times$  6 stimulus$) + ($3 test blocks $\times$  6 stimulus $\times$ 2 repetitions$)]$ = 84 trials.

\subsubsection{Two-Handed Part}
For the two-handed part, we have a 3 $\times$ 2 $\times$ 2 within-subject design with three independent variables: \emph{Input method \{ UM, Once, Swipe \}}, \emph{Device \{ Phone, Tablet \}} and \emph{Orientation \{ Landscape, Portrait \}}.
Participants were instructed to hold the device with their non-preferred hand and were instructed they could use both hands to perform selection.
We did not restrict the specific hand posture, as long as the device was held at least with the non-preferred hand.
Input method was counterbalanced using Latin Square, and Device and Orientation were collapsed into four conditions counterbalanced using Latin Square (Phone-Landscape, Phone-Portrait, Tablet-Landscape, Tablet-Portrait).
A participant would perform 3 input methods $\times$ 2 devices $\times$ 2 orientations $\times [($1 training block $\times$  6 stimulus$) + ($3 test blocks $\times$  6 stimulus $\times$ 2 repetitions$)]$ = 504 trials in the two-handed part.

\subsubsection{Dependent Variables}
We measured accuracy, command selection time.
A trial was considered successful if the participant was able to select the correct target on first attempt. Selection time was measured as the time between the stimulus presentation until the user finished the selection (release of hotkey for UM and Once, end of the slide gesture for Swipe). The modifier key recorded was the last one used before finishing the trial (left or right).
For each individual combination of Technique $\times$ Device $\times$ Orientation, we also measured subjective feedback on a 1-5 Likert Scale (1: strongly disagree, 5: strongly agree) on their overall opinion for the technique ("I liked using this technique"), ease of use, comfort, perceived speed and perceived accuracy.

Each participant took approximately 45 minutes to complete the experiment and were allowed to take breaks between each block. In total, we recorded 12 participants $\times$ 84 trials in one-handed + 504 trials in two-handed = 7056 trials overall.

\subsection{Results}
We conducted two separate statistical analyses for one-handed and two-handed parts.
After gathering data, we removed outliers (data points above average + 3 standard deviations). We then checked the distribution of our data and found it normally distributed.
For the one-handed part, we used paired t-test for accuracy and time (as we only had two Input Methods), and Wilcoxon signed-rank test for subjective measurement.
For the two-handed part, we used three-way ANOVA with repeated measures for accuracy and time, with pairwise t-tests with Bonferroni corrections for post-hoc comparisons.
We applied Greenhouse-Geisser sphericity correction when needed, which corrects both p-values and the reported degrees of freedom.
For subjective measurement, we used Friedman test, with pairwise Wilcoxon signed-rank tests with Bonferroni corrections for post-hoc comparisons.
We do not compare performance between one-handed and two-handed parts, as the design and number of trials differ.
The 95\% confidence intervals shown on the Figures are computed using the z-scores with the following formula: $1.96 \times \frac{SD}{\sqrt{n}}$.

\subsubsection{One-handed Part}
\begin{figure}[!h]
  \includegraphics[width=0.6\columnwidth]{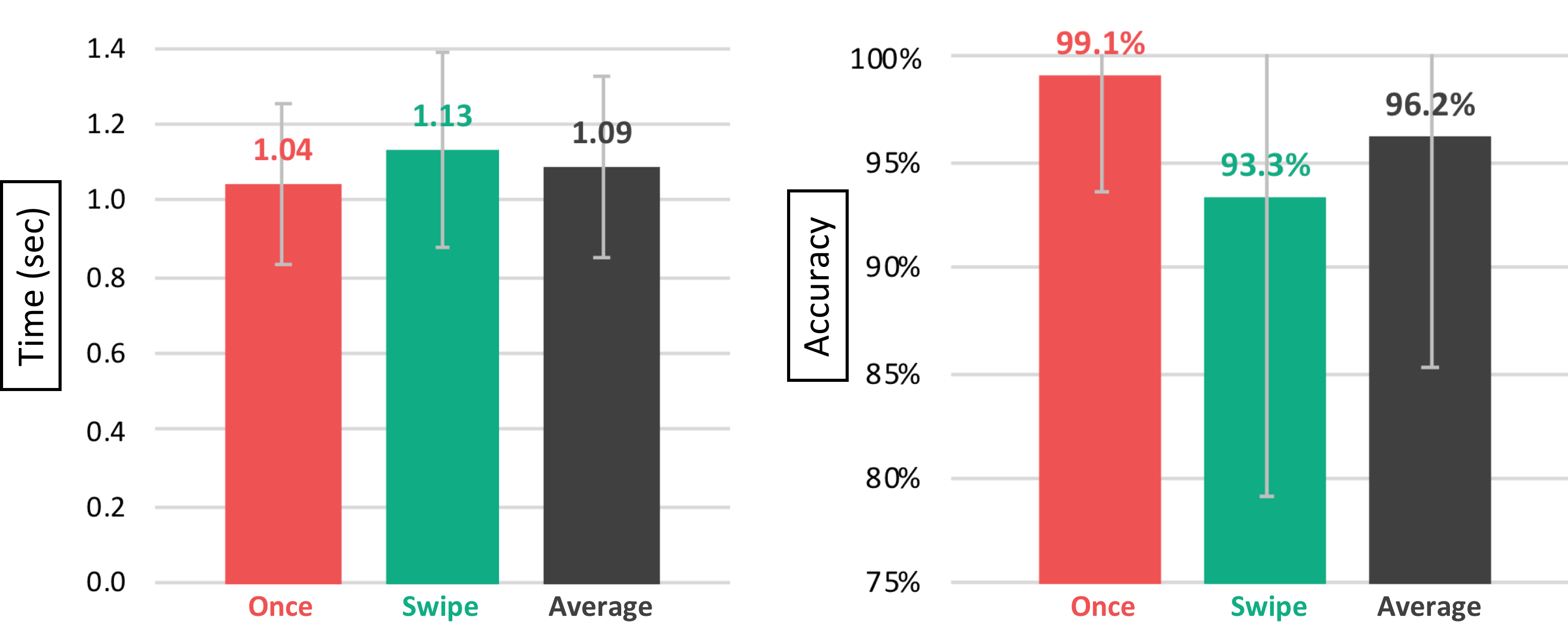}
  \caption{Selection time (Left) and accuracy (Right) performance for the one-handed conditions. Error bars show 95\% confidence intervals. Accuracy ordinate axis starts at 75\%.}~\label{fig:expe2-1handed}
\end{figure}

Participants performed the task rather quickly ($M=1.09s$) and accurately ($M=96.2\%$) as seen in Figure~\ref{fig:expe2-1handed}.
While average command selection time was lower with \emph{Once} ($M=1.04s$) compared to \emph{Swipe} ($M=1.14s$), the difference was not found significant ($p=.10$).
However, we found a significant effect of \emph{Input Method} on accuracy ($t(11)=5.82, p<.001$), as our participants reached higher accuracy with \emph{Once} ($M=99.1\%$) than with \emph{Swipe} ($M=93.3\%$).
In terms of subjective preferences, we did not observe any effect of \emph{Input Methods} on any question, with an overall neutral to slightly positive assessment for both methods ($\frac{3.1}{5}$ to $\frac{3.75}{5}$).

\subsubsection{Two-handed Part}

\emph{Time.}
\begin{figure}[!h]
    \centering
     \includegraphics[width=1.0\columnwidth]{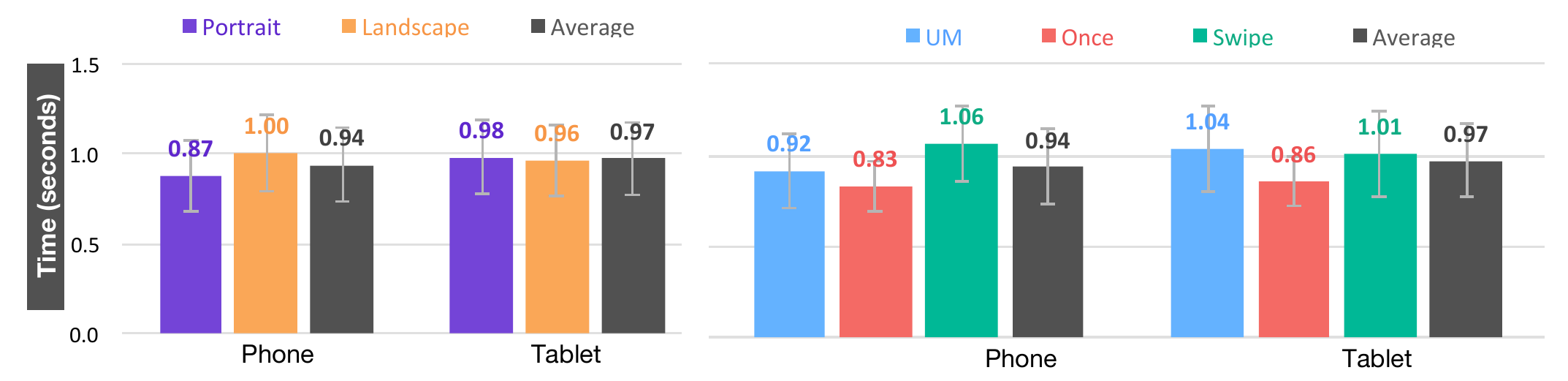}
    \caption{Selection  times (Left) for each orientation and (Right) each input method, on both phone and tablet while using 2 hands. Error bars show 95\% confidence intervals.}~\label{fig:exp2time}
\end{figure}
We found a significant main effect of \emph{Input Method} on time ($F_{2,22}=25.2, p<.0001, \eta^{2}=0.14$). Participants were faster with \emph{Once} ($M=0.85s$) than with Swipe ($M=1.04s, p<.0001$) and \emph{UM} ($M=0.98s, p<.0001$).
We also found a significant main effect of \emph{Orientation} ($F_{1,11}=10.44, p<.001, \eta^{2}=0.02$), with a slightly lower average command selection time in \emph{Portrait} conditions ($M=0.92s$ vs. $M=0.98s$ for \emph{Landscape}).
Performance was very similar on Phone and Tablet, with no effect of \emph{Device} ($p=.18$).
We also found a significant \emph{Device} $\times$ \emph{Orientation} interaction effect on time ($F_{1,11}=23.58, p<.01, \eta^{2}=0.03$), with \emph{Portrait} being overall faster on \emph{Phone} and \emph{Landscape} on \emph{Tablet}.
There was a significant \emph{Device} $\times$ \emph{Input Method} interaction effect on time ($F_{2,22}=7.43, p<.01, \eta^{2}=0.04$), with \emph{Swipe} performing better on \emph{Tablet} than \emph{Phone}, and \emph{UM} following an opposite pattern (Figure~\ref{fig:exp2time}).
We did not find any other interaction effect (all $p>.05$).


\emph{Accuracy.}
\emph{Input Method} had a significant effect on accuracy ($F_{1.15,12.69}=19.16, p<.0001, \eta^{2}=0.25$): our participants were significantly less accurate when using \emph{Swipe} ($M=95.0\%$) compared to \emph{Once} ($M=99.5\%, p<.0001$) and \emph{UM} ($M=99.5\%, p<.0001$).
The average accuracy was significantly higher ($F_{1,11}=8.28, p=.015, \eta^{2}=0.05$) on \emph{Tablet} ($M=98.8\%$) than on the \emph{Phone} ($M=97.1\%$).
We did not find any significant main effect of \emph{Orientation} ($p=.85$) on accuracy.
We also observed a significant \emph{Device} $\times$ \emph{Input Method} interaction ($F_{2,22}=12.36, p<.001, \eta^{2}=0.08$) which is due to the increase of performance for \emph{Swipe} between the \emph{Phone} ($M=92.6\%$) and \emph{Tablet} configurations ($M=97.3\%$).
 \begin{figure}[!h]
   \includegraphics[width=0.6\columnwidth]{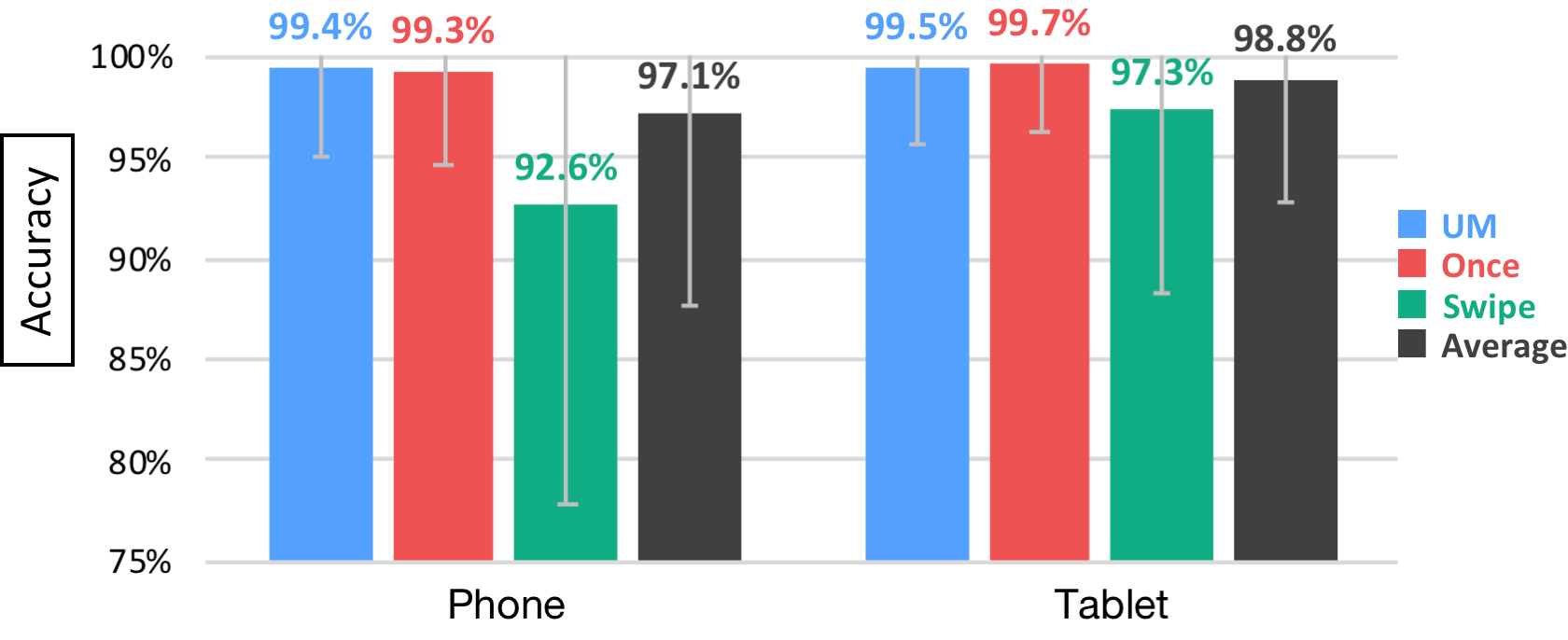}
   \caption{Accuracy performance for individual input method on both phone and tablet for the two-handed conditions. Error bars show 95\% confidence intervals. Ordinate axis starts at 75\%}~\label{fig:expe2-accuracy}
\end{figure}

\emph{Modifier Keys.}
Our participants used the Left modifier key more often. For \emph{Once}, they used it a total of 1103 times (vs. 625 for Right). Usage was similar for \emph{Swipe} (1108 Left vs. 620 Right). The trend is even clearer for \emph{User Maintained} (1265 trials vs. 463 Right). This is likely due to the constrained posture on the left hand, where they could easily access the Left key with their thumbs.

\emph{Subjective Preferences.}
After each individual condition of our experiment, we gathered subjective preferences of our users. We originally analysed the data for each \emph{Device} $\times$ \emph{Orientation} combination, and found out that the results are very similar for each individual condition, and thus decided to aggregate the results (Table~\ref{tab:expe2-quali}). Overall, \emph{Once} received positive scores (scores $> \frac{4}{5}$), while \emph{Swipe} and \emph{UM} received neutral to slightly positive scores.

\begin{table}[!h]
\small
\centering
\begin{tabular}{c|cc|ccc}
\textbf{Question} & \textbf{$\chi^{2}(2)$} & \textbf{$p$-value} & \begin{tabular}[c]{@{}c@{}}\textbf{Score}\\ \textbf{(Once)}\end{tabular} & \begin{tabular}[c]{@{}c@{}}\textbf{Score}\\ \textbf{(UM)}\end{tabular} & \begin{tabular}[c]{@{}c@{}}\textbf{Score}\\ \textbf{(Swipe)}\end{tabular}\\ \hline
Like & 9.52 & $<.01$ & 4.13$^{\alpha}$ & 3.33$^{\alpha}$  & 3.17\\
Ease of Use & 8.77 & $<.05$ & 4.17$^{\alpha}$  & 3.54$^{\alpha}$  & 3.46 \\
Comfort & 6.43 & $<.05$ & 4.13$^{\alpha}$  & 3.46$^{\alpha}$  & 3.46 \\
Speed & 9.91 & $<.01$ & 4.17$^{\alpha}$  & 3.38$^{\alpha}$  & 3.25\\
Accuracy & 8.33 & $<.01$ & 4.42$^{\alpha}$  & 4.33 & 3.25$^{\alpha}$  
\end{tabular}
\caption{\label{tab:expe2-quali}Score (/5) and analysis of subjective preferences. $\alpha$ shows significant ($p<.05$) pairwise differences.}
\end{table}

The results of this experiment are discussed in the final discussion of the paper, along with the results of the next experiment.

\section{Experiment 2: Usage of Input Methods}
The previous study investigated the overall performance of each SoftCuts' input methods under various interaction conditions.
In this study, we want to observe which input method is adopted when the user is not constrained and across different activities like sitting, standing and walking.
Since the 3 input methods are compatible with one another, the users have the freedom to use the one they want depending on their preferences for each specific scenario.
We are interested to see how the results from Experiment 1 would translate when evaluated in a more realistic environment.

\subsection{Participants and Apparatus}
\begin{figure}[!h]
    \centering
     \includegraphics[width=0.6\columnwidth]{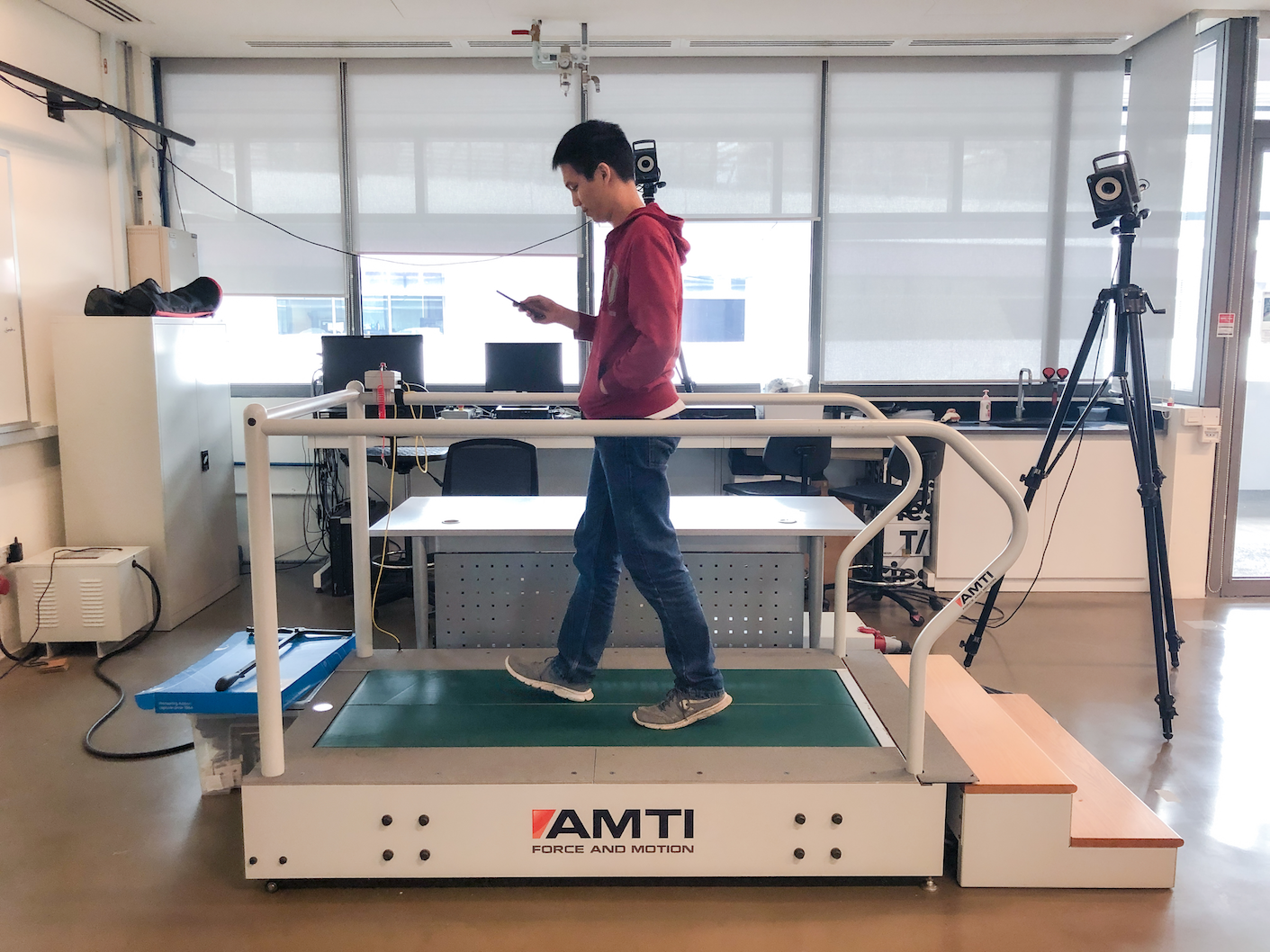}
    \caption{Experimental Setup for the treadmill conditions}~\label{fig:expe5-treadmill}
\end{figure}

Twelve participants (all right-handed, 5 female), aged 21 to 28 ($M=24.2$, $SD=2.72$) were recruited from the university community.
They received the equivalent of 7.5 USD for their participation.
None of these participants took part in the previous experiment.
Only one participant had never used a tablet prior, and we used the same phone and tablet as that in Experiment 1.
In addition, some conditions required the use of a treadmill (see Figure ~\ref{fig:expe5-treadmill}) to simulate a consistent walking speed of 2.5 km/h as in previous works~\cite{Roumen2015,Je2018}.

\subsection{Procedure}
Participants began with completing a questionnaire about their demographic information.
They were then briefed on the experiment and started the practice phase.
During this practice phase, participants were asked to perform 16 command selections with each \emph{Input Method} while seated down.
The command selection was done with the \emph{Phone} and \emph{Tablet}, each \emph{2-handed} and in \emph{Portrait} mode.
The order of presentation of the three \emph{Input Methods} was fully counterbalanced across participants to avoid any bias towards one specific \emph{Input Method}.
The goal of the practice phase was to give participants a chance to familiarize with each input method before starting the experiment.

After practising, participants were briefed on the actual experiment, in which they had to complete three blocks per individual condition (see Design section). Each block has 16 trials (twice each target, in random order).
Participants were instructed to perform the command selection with the \emph{Input Method} they preferred and would rather use in real-life cases.
They were also made aware that they were free to change their choice of \emph{Input Method} during the experiment as they wished. After finishing all conditions, participants were asked to explain their choices of \emph{Input Method} to the experimenter in a short semi-structured interview.

\subsection{Soft Keyboard Layout}
\begin{figure}
    \centering
     \includegraphics[width=1.0\columnwidth]{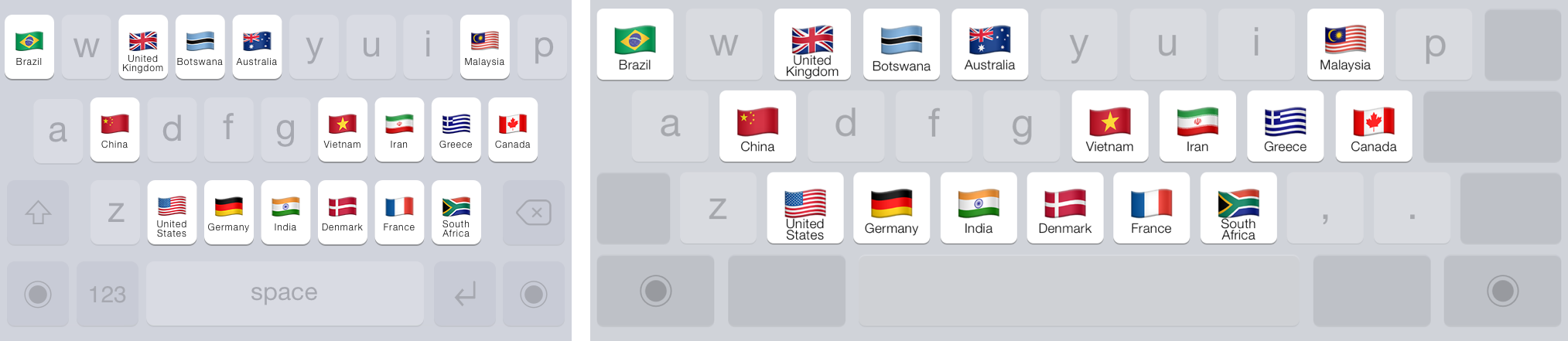}
    \caption{Layout for our second experiment, each shortcut is a country name with its flag. Left: layout on a phone in portrait mode, Right: layout on the tablet in landscape mode.}~\label{fig:expe5-layout}
\end{figure}

Similar to Experiment 1, the soft keyboard used in this study has command keys located at its bottom-left and right corners (Figure~\ref{fig:expe5-layout}).
When either of the command keys was hit, the soft keyboard will be displayed instantly.
Keys with associated commands would be rendered with both its icon and name while keys without associated commands would be greyed out.
This graphical representation is different from both commercial solutions that either only highlight the key \cite{Samsung} or display the name at the top of the key (Microsoft Surface) because we leverage on the ability of icons to deliver efficient guiding cues for visual search \cite{Bailly2016} and conveying command meaning~\cite{Giannisakis2017}.
We chose countries from different continents of the world that have consistent flag shapes, similar to previous studies~\cite{Malacria2013_ExposeHK, Goguey2019}.
The layout for the 16 commands is based on the 16 most common hotkeys on MS Word, which we then mirrored on vertical axis to preserve spatial distribution and have a realistic distribution of commands based on an existing layout.

\subsection{Task and Stimulus}
Each trial begins with the name of a target command displayed on top of the screen. Then, participants would press the modifier key and select the corresponding command key using any of the three available \emph{Input Methods}. Upon selection, either a positive (in green) or negative (in red) feedback would highlight the selected key for 500 milliseconds, before proceeding to the next trial.
For each participant, 8 of the 16 commands were randomly used as stimuli, similar to FastTap~\cite{Gutwin2014}. To ensure similar spatial distribution of selected commands, 2 out of the 8 were from the top row and 3 from each of the middle and bottom row. We also avoided mnemonic associations between keys and countries, e.g. "A" for Australia. 

\subsection{Design}
Our initial design space for this experiment includes: Activity \{Sitting, Standing, Walking\}, Device \{Phone, Tablet\}, Handedness \{1-handed, 2-handed\}, Orientation \{Landscape, Portrait\}.
Complete factorial exploration of this design space would lead to a total of 3 $\times$ 2 $\times$ 2 $\times$ 2 = 24 combinations. Participants were asked to hold both devices similar to Experiment 1. However, to keep the experiment within a manageable time frame, we decided to prioritize some factors depending on the device used, and thus discarded some factors for either \emph{Phone} or \emph{Tablet} conditions. 

For the \emph{Phone} conditions, we deemed \emph{Handedness} as more important than \emph{Orientation}, as one- and two-handed conditions may yield different results. On the other hand, we did not envision a lot of scenarios with physical activity where landscape orientation would be extensively used and thus decided to set the \emph{Orientation} to \emph{Portrait} mode and included \emph{Handedness} in our design.
For the \emph{Tablet} conditions, we assumed that most interactions with a tablet would be two-handed while in motion or simply because supporting the weight with only 1 hand can result in fatigue. As such, we set the \emph{Handedness} to \emph{2-handed} and included \emph{Orientation} in our design.
As such, the \emph{Phone} and \emph{Tablet} conditions are not directly comparable as the impact of \emph{Handedness} and \emph{Orientation} is different. We will thus conduct separate analysis for each device.

In this experiment, we measure the usage of each \emph{Input Method} for our different conditions. For each possible \emph{Input Method}, we count the number of trials where command selection was performed using that specific method. Simultaneously, we also measured the corresponding time taken and accuracy (i.e. whether the right command was selected) for each trial. The order of presentation of \emph{Device}, \emph{Handedness} (Phone) and \emph{Orientation} (Tablet) was fully counterbalanced across participants, while \emph{Activity} was counterbalanced using Latin Square.

Each participant took approximately one hour to complete the experiment. Participants were allowed to take breaks between conditions. In summary, we recorded 12 participants $\times$ 3 activities $\times$ 2 devices $\times$ 2 handedness OR orientation $\times$ 3 blocks $\times$ 8 commands $\times$ 2 repetitions = 6912 trials in total.

\subsection{Statistical Analysis}
The goal of this experiment is to find out which input method participants prefer to use, thus we simply report the usage frequency from our experiment. The results are overall clear, with one technique being chosen most of the time.
We were also interested in finding out whether Activity, Handedness and Orientation have an impact on the usage frequency. To do so, we performed a multinomial logistic regression with R by giving numerical values to our activity levels (1: sitting, 2: standing, 3: walking) and Orientation (1: portrait, 2: landscape). We separated the analysis into three parts: one for Phone 1-Handed (as UM is not usable here), one for Phone 2-Handed, and one for Tablet.

From our results, We then compared the time and accuracy performance. However, given the small number of samples for \emph{Swipe} and \emph{User Maintained}, we only performed statistical analysis for time and accuracy for the \emph{Once} method, using ANOVA.

\subsection{Results}

\subsubsection{Usage Frequency}
\begin{figure}
    \centering
     \includegraphics[width=1.0\columnwidth]{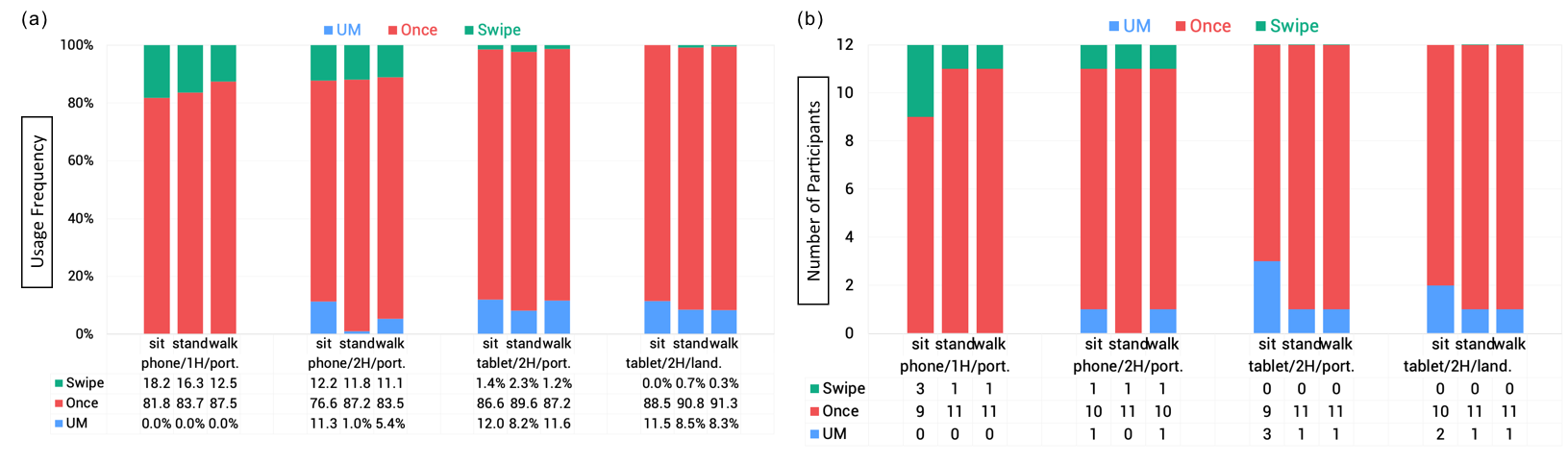}
    \caption{Distribution of each \emph{Input Method} (UM, Once, Swipe) across 4 conditions between device, handedness, orientation while sitting, standing and walking. (a) is derived from actual usage during the experiment while (b) is the declared preference at the end of each condition.}~\label{fig:expe5-usagepref}
\end{figure}

Overall, our users mainly used \emph{Once}, with this method used for 86.2\% of selections. Its usage proportion was at least 76.6\% (Phone 2-Handed Sitting) and at most 91.3\% (Tablet Landscape Walking). The usage frequency of each technique across the different conditions is shown in Figure~\ref{fig:expe5-usagepref}.\\
\\
\emph{Phone 1-Handed.} Participants performed most of the trials using \emph{Once} (84.3\% of selections), and used \emph{Swipe} for the remaining trials (15.7\%). Our analysis shows that \emph{Activity} had an impact on the usage frequency of both techniques (both $p<.001$): the usage of \emph{Swipe} and \emph{Once} varies with \emph{Activity}, from 18.2\%-82.8\% (resp. Swipe and Once) sitting to 16.3\%-83.7\% standing and 12.5\%-87.5\% walking.\\
\\
\emph{Phone 2-Handed.} \emph{Once} was again the most frequently used (82.5\% of selections), with a significant effect of \emph{Activity} ($p<.001$) with usage ranging from 76.6\% while sitting, to 83.5\% while walking and 87.2\% while standing. Swipe was the second most used at a steady 11.7\% of usage, ranging from 11.1\% (sitting) to 12.2\% (walking)  with no effect of \emph{Activity} ($p>.05$).
Finally, \emph{UM} was used for 5.9\% of selections, and was strongly impacted by \emph{Activity} ($p<.001$) with little to no usage while standing (1\%), rather limited while walking (5.9\% usage) and to a level comparable to \emph{Swipe} while sitting (11.3\%).\\
\\
\emph{Tablet.} \emph{Once} was overall used for 89\% of the selections, and its usage was not impacted by \emph{Orientation} or \emph{Activity} (both $p>.05$). \emph{UM} was the second most chosen input method, used for 10\% of the selections consistently across conditions with no visible effect of \emph{Orientation} or \emph{Activity} (both $p>.05$). Swipe was only used for 0.98\% of the trials with no visible effect of \emph{Activity} on its usage. However, we did notice an effect of \emph{Orientation} on \emph{Swipe} usage ($p<.01$): our participants nearly did not use Swipe in Portrait mode (0.3\% of selections) and used it a bit more in Landscape (1.6\% of selections).\\
\\
\emph{Individual Participants Usage.} Among our participants, we did notice that one specific participant (P8) used \emph{Swipe} for 98.9\% of their trials on the \emph{Phone} and \emph{UM} for 99.3\% of their trials on \emph{Tablet}. The participant reported that they did not like the \emph{Once} method overall and that they were used to \emph{Swipe} for typing on soft keyboards on their phone. This suggests that exposure to input methods may have an impact on usage frequency. Our other participants showed a consistent pattern of relying mostly on \emph{Once} for 90\% selections or more and usually trying out the other techniques across conditions.

\subsubsection{Time and Accuracy}
We also analysed the objective performance by filtering trials that were completed using \emph{Once}. We could not perform the analysis for the other two techniques because of the lack of data available for some conditions, as these techniques were seldom used (e.g. \emph{Swipe} on \emph{Tablet} or \emph{UM} on \emph{Phone}).
We performed one analysis for each device because as mentioned in the Design section above, we included \emph{Handedness} for phone's design space and \emph{Orientation} for tablet's.\\
\\
\emph{Phone.} In terms of time, we found significant effect of \emph{Handedness} ($F_{1,10}=9.03, p<.05, \eta^{2}=0.06$) on phone's performance. It took participants on average 1.68s to complete a trial on a phone using 1 hand, and 1.513s using 2 hands. We did not find any effect of \emph{Activity} or interaction (both $p>.05$). For accuracy, we did not observe any effect of \emph{Handedness}, \emph{Activity} or interactions, with an average accuracy of 98.4\%. \\
\\
\emph{Tablet.} For the tablet conditions, there was no significant effect of \emph{Activity}, \emph{Orientation} and no interaction on time (all $p>.05$). A trial would on average last for 1.509s to complete, which is very similar to that of phone's 2-handed performance. In terms of accuracy, there was no significant effect ($p>.05$) of \emph{Activity}, \emph{Orientation} and no interaction, with participants reaching 98.6\% accuracy on average.\\
\\
The time and accuracy results we obtain seem to be in line with Experiment 1, with the performance of \emph{Once} being comparable on Phone-2 handed (0.85s) and Tablet (0.89s). The task in experiment 1 was a simple pointing task, while the task in Experiment 2 required the user to perform a visual search to identify the correct command.
The effect of \emph{Handedness} on time for the phone condition seems consistent with the average performance observed in Experiment 1, where users were overall faster in 2-handed mode (0.85s) vs. 1-handed mode (1.04s). An interesting result from Experiment 2 is that the performance of \emph{Once} does not seem to be affected by \emph{Activity}, making \emph{Once} a robust input method for real life scenarios.

\section{Discussion}
We now discuss our results, with the strengths of each input method and what these results mean for SoftCuts. For each input method, we also discuss how the technique could be adapted to perform multiple hotkey selections sequentially.

\subsection{Is Once the best input method for SoftCuts?}
In our first study, \emph{Once} achieved the overall highest performance in terms of time, accuracy and preference. \emph{Once} significantly stands out, yielding the lowest command selection time both on tablet and phone, regardless of device orientation and they were held.
Participants also achieved near-perfect accuracy of 98-99\% in both experiments.
From experiment 2, we found out that \emph{Once} was chosen as an input method near 4 out of 5 trials, making it the most used input method, no matter the device, handedness or tablet orientation. More importantly, the performance of this method did not seem to be affected by \emph{Activity}, making \emph{Once} the best input method.

Several participants (P1,5,6,7) shared the similar sentiment that \emph{Once} is "versatile" because of its two discrete steps, unlike in \emph{UM} having to communicate between 2 finger actions, which requires more "cognitive effort to use" (P12).
That being said, we must stress that both of our experiments required users to perform a single command selection per trial, which appears to be the best case scenario for this technique.
Indeed, \emph{Once} relies on a pseudo-mode (unlike \emph{UM} which relies on a user-maintained mode). Therefore, it may be less efficient and more tedious to use for performing several command selections at once. This is given that the keyboard would switch back to text entry mode after each command selection, hence requiring users to reactivate command mode for each subsequent command selection.
One way around that would be to allow users to double-tap on the modifier key which would lock the mode to command selection, the same way that one can double-tap the Shift key on a soft keyboard to maintain the upper case mode.

As a very first investigation of SoftCuts, we had to make experimental decisions and trade-offs to keep the length of our studies under acceptable duration.
Future work should investigate the impact of the number of commands to be selected per trial on input method performance and usage.

\subsection{Swipe as a suitable Phone input method}
\emph{Swipe} was a compelling case to investigate.
In experiment 1, we showed that its performance was objectively worse than other input methods. The gap was more significant on the phone, where we observed a lower accuracy overall (around 92\% on average in both 1- and 2-handed), as well as a larger selection time.
The performance of \emph{Swipe} on the tablet was closer to the other techniques, and thus, we expected to see better adoption of \emph{Swipe} on tablets. However, it appears that users simply did not want to use \emph{Swipe} on the tablet, as it was used for around 1\% of the selections, and even less in landscape mode. P4,5,7,8,11,12 found dragging across the larger screen area of the tablet was "challenging" and "uncomfortable". Participants reported that swiping was too "tedious on the tablet", usually because of "the long distance that [one] needs to swipe" (P4).
P8 also reported that swiping allowed them to "change their decision before releasing the gesture at the right key", suggesting that \emph{Swipe} allows user to partially recover from erroneous selections, as long as they did not release their finger.

Despite these issues, Swipe was the second most used input methods for phones, used for 15.7\% of the selections in Phone 1-handed, and 11.7\% of the selections in Phone 2-handed. One participant even exclusively used \emph{Swipe} citing their large usage of swipe based text entry methods and their strong dislike of \emph{Once}. All users did perform part of their selections using \emph{Swipe} on the phone nonetheless. This suggests that \emph{Swipe} could still be the primary input method on the phone for certain categories of users, especially in 1-handed scenarios, despite its objectively worse performance. It is also worth noting that participants tended to rely on \emph{Swipe} more on sitting and standing positions, while the selection was deemed harder while walking.
\emph{Swipe}, by default, does not support multiple selections in a row. It could support that using, for example, a dwell mechanism: users willing to select a specific command would stop on it for a short duration, then proceed to the next command. However, this would likely put the technique at a substantial disadvantage compared to \emph{Once} and even more to \emph{UM}.



\subsection{User Maintained as a Tablet input method}
\emph{UM} is the only of our input methods that require two hands to operate. However, its performance in Experiment 1 is close to \emph{Once} in terms of speed and accuracy, with a small difference of 100 ms for Phone conditions. On the other hand, its subjective preference was overall the lowest, which led us to believe that \emph{UM} may not be used at all.

We were once again surprised to find out that \emph{UM} was used in Experiment 2. In Tablet mode, specifically, it was chosen around 10\% of the time. P5 commented about an ergonomic advantage that they felt with \emph{UM} while supporting the relatively heavy weight of the tablet. They said, "as the thumb holds onto the modifier key at the bottom, a grip action is resembled, hence making it easier and more steady to use the other hand's finger to trigger the (target) key." Another user also performed most of their selection with this method, because of their dislike of \emph{Once}.

Another strong advantage of \emph{UM} over the other techniques is that it readily supports multiple selections, as illustrated in Figure~\ref{fig:expe5-states}. It is important to note that both our experiments featured only a single command selection per trial, putting \emph{UM} in a worst-case scenario. We could thus expect \emph{UM} to be more widely used for productivity tasks, where users select multiple commands at once.

\subsection{SoftCuts in real world}
Both experiments show that users are able to quickly select commands, with an average time around 0.95s in Experiment 1 (pointing only) vs. 1.5s for \emph{Once} in Experiment 2. Accuracy was also extremely high and users enjoyed using the system as seen in the high score of the subjective preferences. This makes SoftCuts a good candidate for command selection technique on both tablet and phones. Our vision for SoftCuts is illustrated in Figure~\ref{fig:banner}.\\
\\
\emph{Input Methods.} A minimal way to implement SoftCuts on mobile devices is to provide \emph{Once} as an input method. However, as highlighted above, some users are likely willing to use \emph{Swipe} on their phone, and \emph{UM} already supports multiple command selection. Based on our state machine presented in Figure~\ref{fig:expe5-states}, the three input methods are compatible with each other, allowing designers to simply implement all three of them without any interference.\\
\\
\emph{Advantages over physical hotkeys.} In addition, SoftCuts make good use of the space used on a soft keyboard. Compared to physical hotkeys, where users need to memorize the mapping between hotkeys and commands, SoftCuts display the name of the command with an icon, making it easier to do command selection even with low familiarity with the mapping of the current application.\\
\\
\emph{Advantages over other command selection techniques.} Compared to other command selection techniques, SoftCuts also potentially leverage users' prior knowledge of hotkeys on desktop computers and can be implemented on any applications, making it a generic and consistent command selection technique across devices. Implementing SoftCuts comes at the small cost of adding a modifier key (either Control or Command) at the bottom left part of the soft keyboard. However, future work is needed to further investigate on the discoverability of SoftCuts.\\
\\
\emph{Keyboard-less scenarios.} While adding one or two modifier key(s) on a soft keyboard layout is rather straightforward for any scenarios where the keyboard is displayed, the same modifier key(s) could be displayed semi-transparently in keyboard-less scenarios (Figure~\ref{fig:banner}). This would partly occlude a small part of the screen, but make the technique extremely salient to the user, as it would lead them to press the key to try SoftCuts. Throughout both experiments, multiple participants suggesting the use of either semi-transparent or fully-transparent modifiers keys for keyboard-less scenarios. The idea of a fully-transparent key was for expert users, as they would remember the overall location of the modifier key on the keyboard layout and would be able to start the selection without the need to look for the exact position of the modifier key. In future work, we would like to evaluate the performance of SoftCuts while carrying a real-world task including keyboard-less applications.

\section{Conclusion}
We explored the concept of soft keyboard shortcuts/hotkeys (SoftCuts) as a generic command selection mechanism for touch-based devices.
We showed that its three input methods yielded high performance, with Once being the best for any device, orientation, handedness, and also most preferred by participants. Interestingly, we found that some participants preferred \emph{Swipe} on phone despite its objectively worse performance on every aspect. As future work, we would like to investigate how to optimize SoftCuts' layout as well as potential skill transfer either from physical keyboards to soft keyboards, or vice versa. 

\bibliographystyle{ACM-Reference-Format}
\bibliography{biblio}


\end{document}